\documentclass[10pt,twocolumn,final]{IEEEtran}

\usepackage{amsmath,amssymb}
\usepackage{graphicx}
\usepackage{algorithm,algorithmic}
\usepackage{tikz}

\newcommand\copyrighttext{%
\footnotesize Copyright 2013 by IEEE. Accepted to IEEE Signal Processing Magazine.}
\newcommand\copyrightnotice{%
   \begin{tikzpicture}[remember picture,overlay]
       \node[anchor=south,yshift=10pt] at (current page.south) {\fbox{\parbox{\dimexpr\textwidth-\fboxsep-\fboxrule\relax}{\copyrighttext}}};
   \end{tikzpicture}%
}

\usepackage{comment}

\newcommand{\field}[1]{\ensuremath{\mathbb{#1}}}

\newcommand{\reals}{\field{R}}
\newcommand{\inputSpace}{\ensuremath{\mathfrak{X}}}
\newcommand{\trp}{{^\top}} 

\newcommand{\divv}{\ensuremath{\mathbb{D}}}

\newcommand{\Ltwo}{\ensuremath{{L}_2}}

\providecommand{\abs}[1]{\lvert#1\rvert}
\newcommand{\norm}[1]{\ensuremath{\left\Vert{#1}\right\Vert}}
\newcommand{\ones}{\ensuremath{\mathbf{1}}}

\newcommand{\vx}{\mathbf{x}} 
 
\newcommand{\vy}{\mathbf{y}} 
\newcommand{\vz}{\mathbf{x}}

\newcommand{\dm}[1]{\,\mathrm{d}#1}
\newcommand{\braket}[2]{\ensuremath{\langle#1|#2\rangle}}

\newcommand{\kernel}{\ensuremath{\kappa}}
\newcommand{\kernelMatrix}{\ensuremath{\mathbf{K}}}

\DeclareMathOperator*{\cov}{cov}

\begin{document}
\title{Kernel methods on spike train space for neuroscience: a tutorial}
\author{Il~Memming~Park, Sohan~Seth, Ant\'onio~R.~C.~Paiva, Lin~Li, and Jos\'e~C.~Pr\'incipe%
\thanks{I.~M.~Park is with the University of Texas at Austin, S.~Seth is with the Helsinki Institute for Information Technology, A.~R.~C.~Paiva was with the Scientific Computing and Imaging Institute at the University of Utah (currently with EM-URC), and L.~Li and J.~C.~Principe are with the University of Florida}
}
\maketitle

\begin{abstract}
Over the last decade several positive definite kernels have been proposed to treat spike trains as objects in Hilbert space.
However, for the most part, such attempts still remain a mere curiosity for both computational neuroscientists and signal processing experts.
This tutorial illustrates why kernel methods can, and have already started to, change the way spike trains are analyzed and processed.
The presentation incorporates simple mathematical analogies and convincing practical examples in an attempt to show the yet unexplored potential of positive definite functions to quantify point processes.
It also provides a detailed overview of the current state of the art and future challenges with the hope of engaging the readers in active participation.
\end{abstract}

\copyrightnotice
\section{Introduction}
\begin{figure}[th!]
\centering
\includegraphics[width=0.48\textwidth]{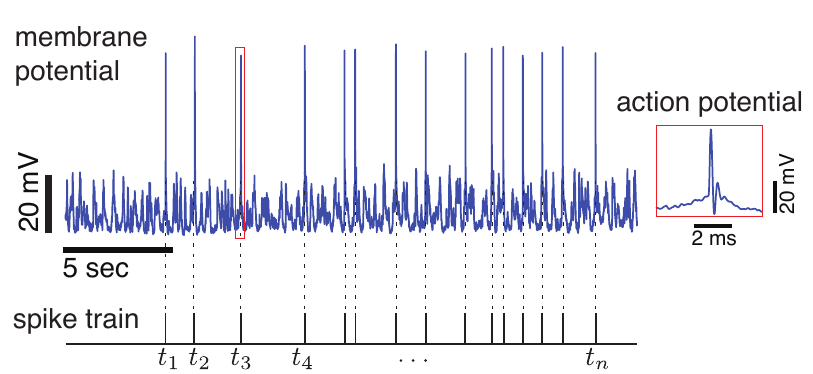}
\caption{
Spike train observation.
Typical intracellular membrane potential recording from a neuron is shown as a continuous time trace.
Each occurrence of an action potential is the only information directly communicated to other neurons through synaptic connections.
Spike train is represented as a sequence of times $(t_1, \ldots, t_n)$ that action potentials are detected.
} 
\label{fig:spiketrain}
\end{figure}

Information processing in the brain is carried out by a complex network of neurons communicating by sending reliable stereotypical electrical pulses known as action potentials, or spikes.
Thus, the information is encoded in a sequence of events over continuous time, and not in the amplitude of the signal as is common in signal processing applications (see Fig.~\ref{fig:spiketrain}).
Studying how information is represented and processed as spike trains---known as the neural coding problem---is one of the key challenges of neuroscience.
We venture to say that the theory of how to represent information in continuous, infinite dimensional spaces is also far from being understood in the signal processing and machine learning communities.
In light of the current signal processing focus in sparseness, point processes (that generate spike trains) are very appealing, since a point process provides the natural limiting case of sparse priors that underlie compressive sensing, and it implements the ultimate sparse representation:
the system only communicates when the information crosses some internal threshold.
This strategy saves power, and provides naturally a sparse representation in time, so the costly step of finding alternative spaces to map the input data for sparseness is unnecessary.
The problem is that the system becomes less observable, and therefore algorithms intended to predict, control or otherwise process the incoming information are less effective and much more cumbersome.
The early attempts in the engineering literature to apply stochastic process theory to zero crossing analysis (a simple way to create a point process) started in the 40's with Rice at the Bell Labs, and found applications in frequency modulation (FM) and shot noise.
The theory of point processes developed primarily in the statistics literature~\cite{Daley1988} and currently this theory is the most widely used approach to quantify spike trains in computational neuroscience as well as in all other areas of science and engineering.
Point processes are also important for machine learning, in particular for online learning that deals with data streams, because of the shortcomings of vector spaces to represent both unbounded data and the resulting inference structure obtained after processing.

Since the spike train space is devoid of an algebra, it imposes many challenges to signal processing methods.
We must then first establish a space for computation or a transformation to a space with the necessary properties.
The approach explained here is to define a proper kernel function on spike trains to capture nonparametrically the temporal structure and the variability of the spike trains of interest.
Once a positive definite kernel is defined, it maps the spike trains into a Hilbert space of functions which allows signal processing tools to be applied directly through the kernel trick.
This methodology has the potential to enlarge the footprint of digital signal processing to objects that are non-numeric, i.e., we can filter spike trains, decompose them in principal components, and perform inference, with exactly the same tools available for time series defined in $\reals$.
But more importantly, the use of kernels provides an opportunity for a myriad of advanced machine learning tools such as Gaussian processes, and probability embedding to be applied to spike trains, opening up a new frontier for next generation spike train signal processing.

\subsection{Neuroscience and neural engineering problems}
\begin{figure}[t]
\centering
\includegraphics[width=0.48\textwidth]{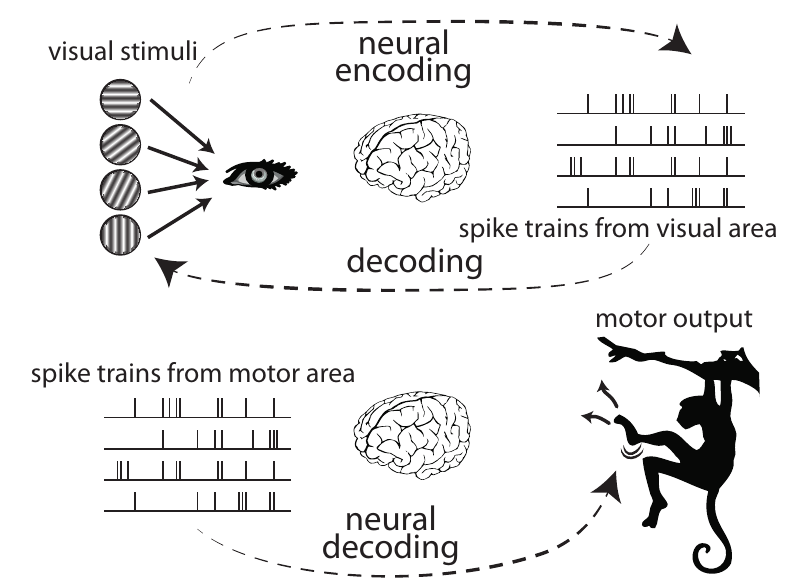}
\caption{
    Illustration of the neural decoding problem in the sensory and motor system.
    Note that decoding is causal for behavior, and anti-causal for sensation and perception.
}
\label{fig:decoding}
\end{figure}
The idea of a neural code is prevalent in the sensory and motor systems where the variables of interest are directly observable, although it is latent in all neuronal communication.
In a sensory system, we would often like to understand how the sensory stimuli are encoded and transformed in each stage of neural processing.
For example, visual stimuli excite a cascade of neurons in the visual pathway from photoreceptor neurons, and retinal ganglion cells in the eye to various areas of the visual cortex.
By analyzing the spike trains, we can understand how certain aspects of a stimulus are processed and represented in these neurons~\cite{Olshausen2005}.
The study of neural code often consists of 
\begin{enumerate}
    \item identifying neurons that encode certain features of interest (neuron identification), and
    \item finding the functional relation between the feature and spike trains of the identified neurons (neural encoding/decoding).
\end{enumerate}

\begin{figure}[t!h]
\centering
\includegraphics[width=0.48\textwidth]{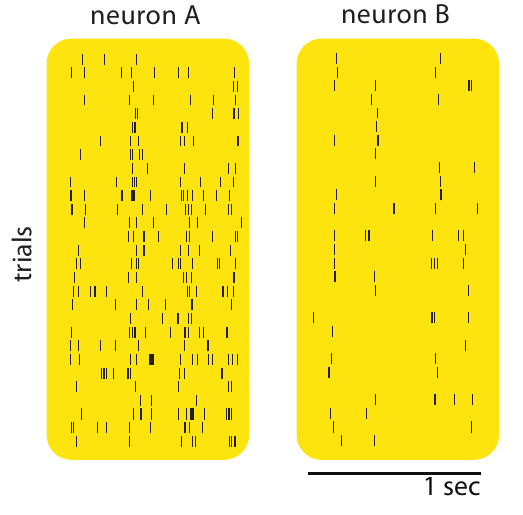}
\caption{
\textbf{Trial-to-trial variability.}
Neural response to repeated stimuli are variable.
The variability structure can differ for different neurons and contexts.
These trials are from retinal ganglion neurons~\cite{Jacobs2006}.
}
\label{fig:ttv}
\end{figure}

A major challenge in neuron identification is the neural variability, or ``noise'', in the system.
For example, when a fixed stimulus is repeatedly presented to the system, the trial-to-trial variability of the neural response is often complex (see Fig.~\ref{fig:ttv}).
Therefore, to determine if a neuron encodes for the variable of interest $x$, one cannot compare simply single trial responses, but requires \textit{collections} of responses that are samples from the stimulus-conditional distribution $p(\mbox{spike train}|x)$.
Fortunately, the tools from signal detection theory and hypothesis testing can be extended to stochastic spike trains via kernels to solve the neural identification problem (section~\ref{sec:mmd}).

On the other hand, a better method for reading out the stimulus from the spike trains (neural decoding) can have major impact on a number of clinical and biomedical applications.
For example, it can improve sensory prosthetics such as cochlear implants, which are widely used, and retinal and tactile prosthetics that are under active development~\cite{Wilson1991,Humayun2003,Chapin2004}.
In motor systems, identifying which neurons are involved in motor planning and control, and understanding how information is represented in spike trains is essential in building motor prosthetics~\cite{Nicolelis2009,Chapin1999}.
Similar approaches have been taken for various higher-level cognitive systems such as decision making, memory, and language~\cite{Berger2001}.

Spike train kernels provide alternative tools to the neural coding problem (Fig.~\ref{fig:decoding}).
Traditionally, the ``rate code'' has been the dominant idea in neuroscience~\cite{Adrian1926b,Hubel1959}, and it has been repeatedly demonstrated experimentally.
The rate code hypothesis states that the average spike count encodes all information underlying the stimulus, i.e., that the spike timing is not useful for neural processing.
Contrary to the rate code hypothesis, there is also ample evidence for the so-called ``temporal code'' hypothesis which states that extra information is encoded in spike timings~\cite{Dayan2001,Sejnowski1995,Hopfield1995}.
The neuroscience community, however, has largely relegated the possibility of a temporal code to a secondary role perhaps due to the large dimensionality of the neural code space and the limited ability of statistical methods that directly operate on spike trains and are powerful enough to discover new patterns.
If the brain processes and communicates sensory data optimally amongst neurons, one natural solution is to utilize a representation that preserves as much of the information as possible~\cite{Barlow1959}.
Along this line of reasoning, the timing hypothesis should be the preferred theory because it is the one that guarantees no loss of information and it solves the conundrum: in cases where it is impossible to use rates (when the response time has to be minimized), spike times are preferred, but a representation that is sensitive to spike times also can easily represent rates by integration.
Practically the argument between rates and timing is also biased by the degree of sophistication of the mathematical tools required:
it is difficult to quantify spike timings, while it is very easy to process rates, therefore there may have been many experimental observations that corroborate the spike timing hypothesis that were never published because researchers could not quantify appropriately their data.  
Spike train kernels shine a new light into this controversy by providing a general framework for studying spike trains that can accommodate both hypotheses.
We hope that, by focusing on what is common, the spike train kernel approach may kindle experimental research to show that different neuron classes are optimized for different time scales of processing, just like engineers design differently transistors for high speed CMOS and sample and hold circuits.  

\subsection{Kernels and kernel methods}

The practicality of signal processing is due to a clever exploitation of the linear model.
Unfortunately, not all the problems we want to solve are well approximated by the linear model.
The Hilbert space approach~\cite{Mate1989}, and more specifically the reproducing kernel Hilbert space (RKHS)~\cite{Scholkopf2002} extend the linear model mathematics to nonlinear modeling in the input space.
The methodology is principled because it provides a general way to handle different types of nonlinearity, the optimization is convex, and the methodology is still practical in terms of computational complexity.
But in our opinion, the true importance of kernel methods for neural signal processing is their ability of map abstract objects to an Hilbert space---a linear functional space equipped with an inner product.
Indeed, at the core of the above mentioned problems in neuroscience to quantify spike trains is the lack of standard algebraic operations such as linear projection and linear combination for spike trains.
This mapping supplies the required structure for applying most signal processing tools, and also allows otherwise complex nonlinear computation.

The theory of reproducing kernel Hilbert spaces provides a foundation for the existence of a (possibly) infinite dimensional Hilbert space---a feature space---associated with any positive definite function of two arguments called a kernel~\cite{Aronszajn1950,Scholkopf2002}.
Let the input space data $\inputSpace$ be an object---e.g., a point in $\mathbb{R}^3$, a graph, or a spike train---and a kernel $\kernel: \inputSpace \times \inputSpace \rightarrow \reals$ be a real-valued bivariate function defined in the input space $\inputSpace$.
The input sample $x \in \inputSpace$ is mapped to the RKHS as the function $\kernel(x, \cdot)$, therefore the kernel specifies the richness of the transformation.
The kernel defines also the inner product of the Hilbert space, i.e. the kernel $\kernel(x,y)$ provides the similarity in the RKHS of the functional images of any two samples $x$ and $y$ in the input space and encapsulates any prior knowledge about the input space structure.
Moreover, the inner product of two functions in the RKHS can be computed by a scalar kernel evaluation in the input space, i.e. $\braket{x}{y}_\mathcal{H} = \kernel(x,y)$.
This property brings computational simplicity, therefore we have a principled framework that allows nonlinear signal processing with linear algorithms and makes working with functions practical.

For any finite set of points in the input space $\{x_i\}_{i=1}^n$, the resulting matrix $\kernelMatrix$,
\begin{align}\label{eq:def:kernel:matrix}
    \kernelMatrix = 
    \begin{bmatrix} 
	\kernel(x_1,x_1) & \kernel(x_1,x_2) & \dots & \kernel(x_1,x_n)\\
	\kernel(x_2,x_1) & \kernel(x_2,x_2) & \dots & \kernel(x_2,x_n)\\
	\vdots & \vdots & \ddots & \vdots\\
	\kernel(x_n,x_1) & \kernel(x_n,x_2) & \dots & \kernel(x_n,x_n)
    \end{bmatrix}
\end{align}
must be symmetric and positive semi-definite for a proper kernel $\kernel$, i.e., for any real vector $\vz \in \reals^n$, $\vz\trp \kernelMatrix \vz \geq 0$.
Given data in $\inputSpace$, the kernel matrix $\kernelMatrix$ represents the inner product between each pair in the Hilbert space. 
The kernel matrix plays a central role in kernel method algorithms, because for most algorithms, it contains all information required about the input.

Let us illustrate the importance of the kernel design with kernel mappings on the real line where we know the feature space explicitly.
If we map $x \in \reals$ to a three-dimensional feature vector $[1, \sqrt{2}x, x^2] \in \reals^3$, then linear regression in the feature space corresponds to a quadratic fit in $\reals$.
Equivalently, this quadratic fit can be achieved by kernel least squares using the polynomial kernel $\kernel(x,y) = (1 + xy)^2$ without explicitly constructing the feature space~\cite{Scholkopf2002}.
This is because the least squares linear regression only requires operations provided by the Hilbert space (linear combination, and inner product) and the polynomial kernel is the inner product of the feature space.
The advantage of kernel method is avoiding the intermediate feature space representation especially when it is of high dimension.

One popular kernel is the Gaussian (a.k.a. squared exponential) kernel 
$\kernel(x,y) = \exp(-(x-y)^2/\sigma)$,
which implicitly corresponds to an infinite dimensional feature space.
It captures the local similarity in the real line; $x$ and $x + \epsilon$ are assumed to be very similar if $\abs{\epsilon} \ll \sigma$, and gradually becomes dissimilar as $\abs{\epsilon}$ increases.
However, the choice of $\sigma$ is critical.
As the kernel size parameter $\sigma$ tends to zero, it approaches the trivial kernel, $\kernel(x,x) = 1$, and $\kernel(x,y) = 0$ for $x \neq y$, that still maps the input to an infinite dimensional space where every mapped input point is orthogonal to each other, and hence the feature space has no ability to generalize and basically acts as a lookup table.
On the other hand, if $\sigma$ is larger than the dynamic range of the data, the Gaussian kernel provides basically a constant mapping of the argument, therefore the feature space is unable to weight distances differently and looses the ability to discriminate distinct points.
These two example kernels are extremes that do not bring any advantage.
In practice, we use a kernel (and kernel size) that is in between, one that provides a rich feature space with proper smoothing such that the practitioner can use it to nonlinearly interpolate between data points (e.g., in the scale of $\sigma$ for the Gaussian kernel).

\section{Spike train kernels}
In the previous section we discussed kernel methods on the real line, but the beauty of the theory is that it can be applied to other more abstract spaces.
In fact, various discrete objects naturally represented as a collection such as graphs, sets, and strings have been successfully embedded in Hilbert spaces through kernels~\cite{Cortes2004,Haussler1999,Shin2008}, opening the door for many traditional tools to be directly applied when a suitable kernel is used.
The only difference with respect to the Gaussian kernel is that we now need a way to define the kernel in a way that is relevant to measuring the similarity between spike trains.
Once this is done, we can replicate the same operations as explained for the Gaussian kernel on the real line, i.e., we can quantify similarity between spike trains and define a space where we can build signal processing models and perform inferences.   

In the remainder of the section, we introduce several important spike train kernels and discuss their advantages and disadvantages.

\subsection{Count and Binned Kernels}
A trivial way of constructing a spike train kernel is by first mapping the spike trains into a finite and fixed dimensional Euclidean space and using its inner product: $\vx, \vy \in \reals^d, \kernel(\vx, \vy) = \vx\trp\vy$.
For example, simply counting the total number of spikes in each spike train maps spike trains into natural numbers which is a subset of the real line.
The resulting kernel is called the \textit{count kernel}.
Count kernel completely ignores the timing, but it is useful because it encompasses the conventional rate coding analysis done by neuroscientists.
For instance, the optimal least squares linear decoder is equivalent to the kernel least squares, and the test of mean rate difference is equivalent to MMD (see section~\ref{sec:mmd}) with the count kernel.

\begin{figure}[t!h]
    \centering
    \includegraphics[width=0.48\textwidth]{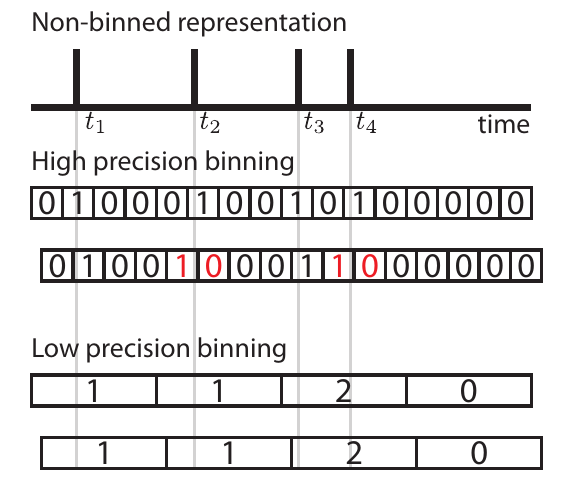}
\caption{
    Advantages and disadvantages of fine and coarse time binning representation.
    The continuous time axis is divided into a sequence of fixed intervals,
    and the number of spikes occurring within each time bin is counted.
    Two bin sizes with different offsets are shown.
    Red numbers indicate changes in the representation due to the small temporal jitter.
    Small bin sizes allow representation of fine temporal structure,
    while small fluctuations causes radical changes if the spike trains
    are considered as a vector in the Euclidean space.
    Larger bin sizes smooth the time, so they are less sensitive to
    small fluctuations but detailed temporal information is lost.
    Hard boundaries of binning can be relaxed by using a smooth basis function centered at fixed times, such as a Gaussian or raised cosine function instead of a rectangular window.
    However, the resulting feature space still has limited dimensionality because of the quantization imposed on time by these techniques.
}
\label{fig:binning}
\end{figure}

A na\"ive extension of the count kernel is to bin the spike train by choosing a sequence of consecutive time windows (Fig.~\ref{fig:binning}).
Neighboring time bins corresponds to different dimensions, and hence the kernel does not bring any smoothing across the bin boundaries.
In the limit of fine time binning, all information of the continuous representation is preserved in the binned representation at the expense of huge dimensionality.
When combined with the Euclidean inner product, binning in this regime is catastrophic because the inner product implements a look-up table, like the trivial kernel mentioned earlier.
On the other hand, when the bin size is larger, the temporal continuity within each time bin is respected, and some smoothing is provided, however, the resulting feature space is low dimensional, and temporal details in the spike trains cannot be fully represented.
For some applications, there is a sweet spot for the bin size that may perform well, since it can partially extract linear relations with respect to the rate code hypothesis~\cite{Serruya2002}.
The linear model on binned data, so popular in brain machine interfaces (BMIs)~\cite{Nicolelis2009} is one example of this technique.

\subsection{Spikernel}
Although directly binning spike trains as objects in the Euclidean space can be misleading, a better kernel can be constructed using this representation but different inner product.
The first successful kernel for neuroscience is the \textit{spikernel}~\cite{Shpigelman2005} which falls in this category.
It allows local time manipulation (time warping) enabling spike counts in neighboring bins to be matched, effectively smoothing over bin boundaries.
It also weights different time bins according to the distance from the time of interest, which is a reasonable assumption based on the finite memory of neural systems.
The spikernel has been successfully demonstrated to perform better than binned count kernel in the context of brain-machine interfaces (decoding the motor signal from neural activity)~\cite{Shpigelman2005}.

In general, the spikernel performs robustly~\cite{Park2012a}, however, it fundamentally lacks the ability to control its temporal precision since it is tied to a binned representation.
In addition, it should be noted that the spikernel is computationally expensive to evaluate, and it requires tuning five free parameters, including bin size. 
The relatively large number of free parameters delivers a flexible kernel, which is supported by its performance, but tuning these parameters requires an extensive optimization that hinders its appeal.

\subsection{Linear functional kernels}
\label{sec:linear}
As we have seen earlier, the binning transformation is lossy---many spike trains can be mapped to the same binned representation---and similar spike trains can be mapped to quite different representations.
How can we avoid binning and preserve all information and create a positive definite kernel?
One solution is to use an infinite dimensional representation~\cite{Paiva2008b}.

\begin{figure}[t!]
    \centering
    \includegraphics[width=0.48\textwidth]{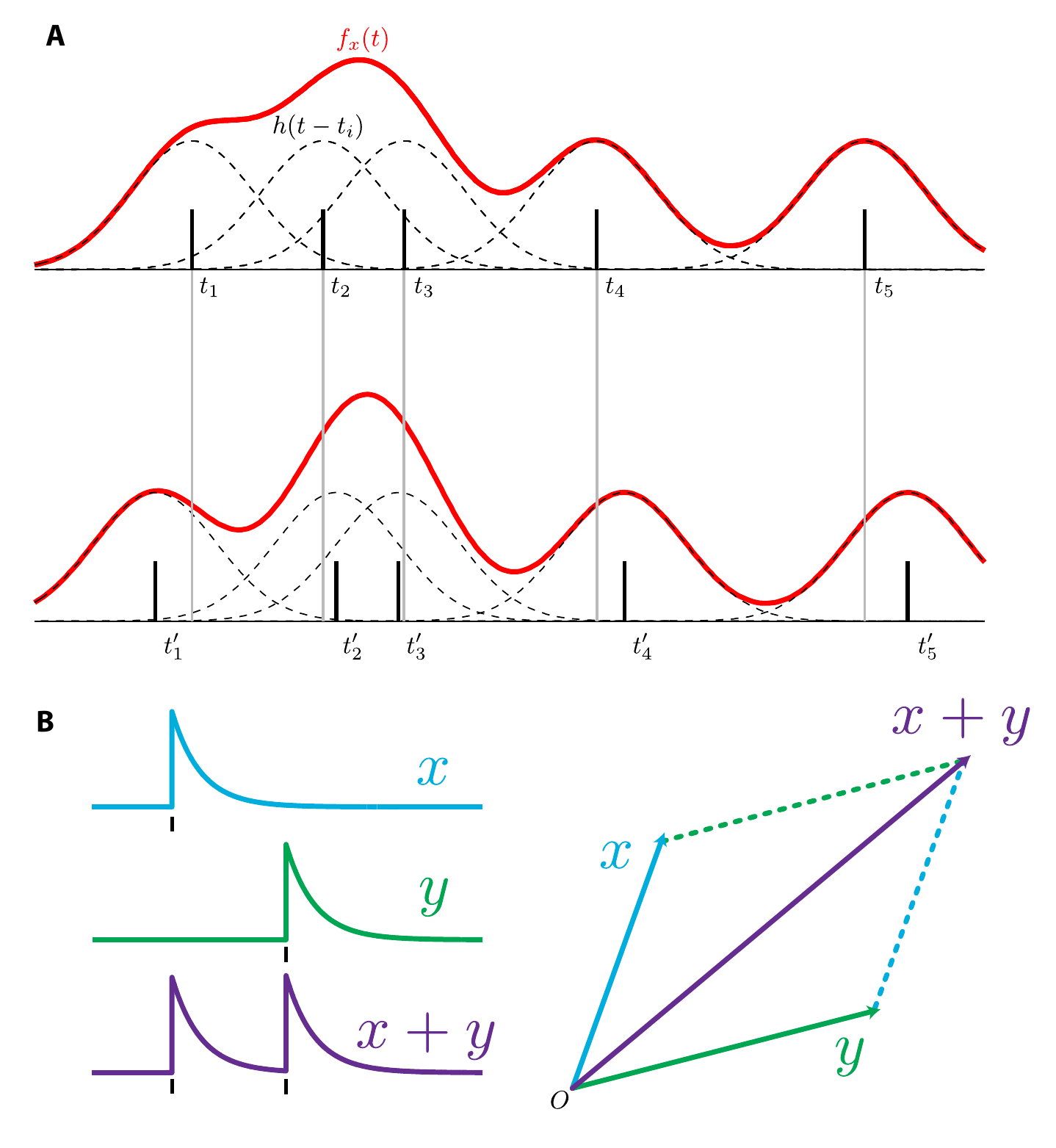}
    \caption{Linear functional kernel.
	(A) Two spike trains smoothed with a function $h(t) = e^{-(t/\sigma)^2}$ in red.
	The spike timings of the first spike train are jittered to generate the second spike train, and their smoothed representation are similar in $\Ltwo$.
	(B) Demonstration of linearity of linear functional kernels using $h(t) = e^{-t/\tau} I(t \geq 0)$.
	The third spike train is a superposition of the first two spike trains, and so is the corresponding smoothed representation.
	A schematic of the three vectors in $\Ltwo$ explicitly indicates their linearity.
    }
    \label{fig:linear}
\end{figure}

Let $h$ be a finite energy impulse response of a linear filter over time (possibly non-causal), and represent the spike train as a sum of Dirac delta functions $x(t) = \sum_i \delta(t - t_i)$ (non-binned representation, see Fig.~\ref{fig:binning}).
Each spike train can be uniquely transformed into a function via convolution with a non-trivial $h$:
\[
    f_x(t) = x \ast h = \sum_i h(t - t_i).
\]
The resulting transformed spike train $f_x(t)$ is a function in $\Ltwo$ space, which is a Hilbert space on its own.
The inner product in $\Ltwo$ is defined as,
\[
    \braket{f}{g} = \int f(t) g(t) \dm{t}.
\]
By choosing a locally concentrated $h$, spike trains with similar small jitter are mapped to similar functions (Fig.~\ref{fig:linear}A).
Therefore, it is continuous with respect to small temporal perturbations.
The \textit{linear functional kernel} is simply defined as,
\begin{align}\label{eq:linear:def}
    \kernel(x,y) = \braket{f_x}{f_y} = \int (x \ast h)(t) (y \ast h)(t) \dm{t},
\end{align}
that is, the inner product of the two smoothed functional representations in $\Ltwo$ (Fig.~\ref{fig:linear}).
Note that \eqref{eq:linear:def} can be rewritten with an explicit summation over all pairs of spikes,
\begin{align*}
    \kernel(x,y) = \sum_{i,j} \int h(t - t^x_i) h(t - t^y_j) \dm{t}
    	= \sum_{i,j} g(t^x_i, t^y_j)
\end{align*}
where $g(u,v) = \int h(t - u) h(t - v) \dm{t}$.
Therefore, the kernel evaluation in the infinite dimensional space can be computed with $O(N_x N_y)$ function evaluations of $g$ where $N_x$ is the number of spikes in the spike train $x$.

A few special choices of $h$ are worth noting~\cite{Schrauwen2007,Paiva2008b}.
If the smoothing function has the form of a Gaussian function, 
$h(t) = e^{-t^2/2\sigma^2}$, then 
$g(u,v) = e^{-(u-v)^2/\sigma^2}$.
More importantly, if the smoothing filter is a causal exponential decay $h = e^{-t/\tau}$ for $t > 0$, then we obtain the following kernel:
\begin{align}\label{eq:mci}
    \kernel(x,y) = 
    \sum_{i,j} \exp\left(- \frac{1}{\tau}\vert t^x_i - t^y_j \vert \right).
\end{align}
This spike train kernel has two distinct advantages. 
First, it has a neurophysiological interpretation, since the synaptic transfer function that transforms the spike trains to an analogue intracellular signal in the downstream neuron can be approximated as a first order dynamical system (i.e., first order IIR filter)~\cite{vanRossum2001,Paiva2010b}.
Second, it can be computed in $O((N_x+N_y) \log(N_x+N_y))$ time~\cite{Paiva2008b}.
The fast computation is due to properties of the double exponential function~\cite{Chen2006a}.

It is easy to see that the linear functional kernels are linear with respect to the superposition of spike trains.
If $x(t)$, $y(t)$, and $z(t)$ are spike trains represented as sum of delta functions, $\braket{x+y}{z} = \braket{x}{z} + \braket{y}{z}$ (Fig.~\ref{fig:linear}B).
Therefore, the functions built on this space have the same constraint;
the function value for the superposition $x(t) + y(t)$ is the sum of the function value for $x(t)$ and $y(t)$.
Note that the binned spike train kernels share this property of linearity with respect to superposition.
We will see that this limitation can be a critical weakness.

\subsection{Nonlinear functional kernels}

To unleash the full potential of kernel methods, we need binless nonlinear spike train kernels.
There are several ways to extend the linear functional kernels to be nonlinear~\cite{Park2012a,Park2010b,Paiva2008b,Paiva2010b,Houghton2009a}.
Here, we focus on building the Schoenberg kernel since it provides a provably universal kernel.
Schoenberg kernels are derived from the radial basis functions, and takes the following form,
\begin{align}\label{eq:shoenberg:form}
    \kernel(x,y) &= 
	\phi \left(
	    \norm{x-y}^2
	\right)
\end{align}
where the function $\phi: [0, \infty) \rightarrow \reals$ is completely monotone on $[0, \infty)$ but not a constant function~\cite{Schaback2000}.
Examples of completely monotone functions are $e^{-\alpha x}, \frac{1}{(x+\alpha^2)^\beta}$ where $\alpha$ and $\beta$ are constants~\cite{Berg1984}.

We take the functional norm derived from the linear functional kernel, that is,
\begin{align*}
    \norm{x-y}^2 &= 
	\braket{x-y}{x-y} = \braket{x}{x} - 2\braket{x}{y} + \braket{y}{y}.
\end{align*}
Next, we build radial basis kernels on top of the feature space induced by the linear functional kernel.
Therefore, 
in Schoenberg kernels, the underlying linear functional kernel provides the smoothness in the space, and the radial basis function $\phi$ enforces the linear superposition to only hold locally.
This combination guarantees the resulting kernel to be powerful for both neural identification and decoding applications.

A typical choice is to use \eqref{eq:mci} as $\kernel'$ with $\phi(x) = e^{-\alpha x}$ which results in the following form:
\begin{align}\label{eq:schoenberg}
    \kernel(x,y) = \exp\left(-\frac{1}{\sigma^2}\left(\kernel'(x,x) - 2\kernel'(x,y) + \kernel'(y,y)\right)\right).
\end{align}
This can be considered as an analogue of the widely used Gaussian kernel for Euclidean space.
Schoenberg kernels are universal, in the sense that they can asymptotically approximate arbitrary nonlinear function from spike trains to reals.
They have an additional scale parameter $\sigma$ which controls how much smoothing is applied to the space induced by the base kernel.

\subsection{Extending single spike train kernels to multiple neurons}
So far we have introduced kernels that compute similarity between a pair of spike trains (either from a single neuron at different times or from a pair of neurons).
However, recent recording techniques allow simultaneous recording of up to a couple of hundreds of neurons.
Thus, we need kernels for a pair of \textit{sets of} spike trains from many neurons.
There are a couple of simple yet effective ways to extend single neuron kernels to multiple neuron kernels (see~\cite{Genton2001,Scholkopf2002} for combining kernels).
First is to use a product kernel,
\begin{align}\label{eq:product}
    \kernel(x,y) = \prod_i \kernel_i(x_i, y_i)
\end{align}
where $i$ indexes over simultaneously recorded neurons.
Second is to use a direct sum kernel,
\begin{align}\label{eq:direct:sum}
    \kernel(x,y) = \sum_i a_i \kernel_i(x_i, y_i),
\end{align}
where $a_i$ are weights for combining the effect of each neuron.
The product kernel is the natural inner product of the product Hilbert space, and the direct sum kernel is that of the direct sum Hilbert space.
The product kernel preserves the universality of elementary kernels (e.g., with Schoenberg kernels), but if the effective dimension of the spike train manifold increases (as in the case of less dependent spike trains and/or independent noise processes) the number of spike trains required to ``fill'' the space increases for the same kernel size.
Hence, more smoothness may have to be incorporated (imposed by kernel sizes), or exponentially more data may be required to estimate equivalently detailed nonlinear functions.
The direct sum kernel does not preserve universality; in fact, only additive functions over multiple neurons are spanned by those kernels.
Therefore, unless such constrains are of interest, it is not useful for general neuroscience applications.
In general, combining kernels increases the number of hyperparameters, making cross-validation less practical, hence we recommend empirical Bayes methods for their estimation~\cite{Rasmussen2005,Bishop2006}.

Although it is possible to form a product kernel from the spikernel, it is not necessary to do so because the spikernel can be extended directly for multiple neurons by considering a vector of spike counts for each time bin~\cite{Shpigelman2005}.
In such construction, the time warping is uniformly applied to all spike trains.
Since the time complexity is only additive for the number of neurons, for a large population recording, spikernel could be computationally advantageous.

\section{Applications}
Equipped with spike train kernels, we can now discuss application areas in neuroscience and neural engineering, each of which requires a different class of kernel methods.
We discuss the problem of hypothesis testing first, followed by stationary neural code analysis using regression and online neural decoding with adaptive filtering.

\subsection{Neuron identification}\label{sec:mmd}
Due to the trial-to-trial variability of neural responses (Fig.~\ref{fig:ttv}) the collection of responses to repeated stimuli can be considered as realizations of a random process.
When realizations are spike trains, the corresponding mathematical object, the probability law over spike trains, is called a \textit{point process}.
It is often necessary to determine if two sets of responses given different experimental conditions are different---we want to know if the response carries any information about the experimental condition of interest.
For example, some neurons in the visual cortex encode the stimulus color regardless of the motion, while some encode the directional motion regardless of the color.

In practice, a severe bias may be unwillingly included when searching for a neuron that encodes information about a certain feature especially in the context of \textit{in~vivo} electrophysiology.
In a typical setting, a trained electrophysiologist would listen to the firing pattern of each neuron and make a decision on the fly to record from the probed neuron, which tends to be the one with larger firing rate modulation.
\begin{figure*}[t!h!]
    \centering
    \includegraphics[width=0.95\textwidth]{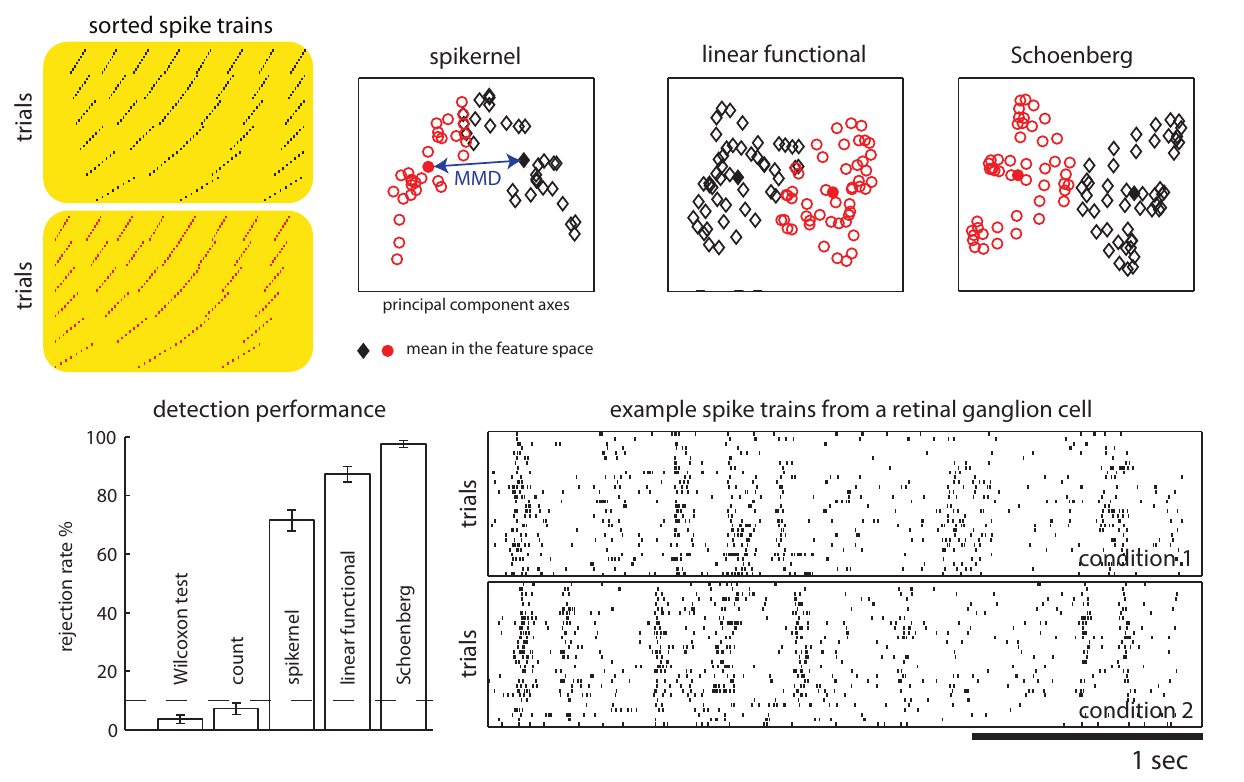}
    \caption{
	Illustration of maximum mean discrepancy (MMD) and hypothesis testing results.
	\textbf{(Top row)}
	Two sets of artificial spike trains shifted half a cycle is analyzed with three different kernels: spikernel, linear functional kernel~\eqref{eq:mci}, and Schoenberg kernel~\eqref{eq:schoenberg}.
	The spike trains mapped in the feature space are visualized in two dimensional subspace determined by the kernel principal component analysis that preserves the most of MMD~\cite{Park2012a}.
	The distance between the mean in the feature space corresponds to MMD.
	In this example, all three kernels successfully rejected the null hypothesis, indicating the corresponding point processes are distinct.
	\textbf{(Bottom row)}
	Hypothesis testing performance on spike trains recorded from retinal ganglion neurons.
	We used the rat retinal ganglion cell recordings from Nirenberg's lab~\cite{Jacobs2006} (22 stimulus conditions repeated 30 times for 3.1 sec from 15 neurons).
The task is to discriminate different stimulus conditions given the spike trains from a single neuron.
	For each neuron, we paired the stimulus conditions to have close median firing rate response by sorting the conditions and pairing neighbors for the test.
	Higher rejection rate implies a more practical divergence for the test.
	Dashed line is the test size ($p = 0.1$).
	Error bar indicate the standard error (165 tests).
	(Copyright 2012 MIT Press Journals. Modified from~\cite{Park2012a} with permission).
    }
    \label{fig:divergence}
\end{figure*}
Identifying neural selectivity has been widely done assuming that the information is only represented in the mean statistics (firing rate).
The conventional estimator is a histogram (or a smoothed version), known as the peri-stimulus time histogram (PSTH), which is obtained by binning the time, and averaging over trials.
The difference in the mean statistics is then used for associated hypothesis testing~\cite{Eggermont1983,Naud2011}.

Another widely used test is the Wilcoxon rank sum test on the spike counts~\cite{Kreiman2000}.
It nonparametrically captures the difference in the distribution over counts, therefore, more than just the mean count is tested.
Since the count distribution must span across multiple values to be meaningful, it requires a large window that captures relatively many spikes, and it is difficult to apply it to multiple bins.
Thus, the count distribution cannot capture the differences in the temporal structure.
For these reasons, these widely used parametric statistical tests are fundamentally incapable of discriminating many features.

Instead, we submit that a class of nonparametric statistics is needed that can discriminate among many point process distributions, either in terms of higher order  statistics or the temporal dimension.
Such a statistic is known generally as a \textit{divergence}~\cite{Seth2010a,Park2010b,Park2011d,Park2012a}.
One can define a divergence measure from kernels by utilizing a recent development  known as \textit{probability embedding}, which provides a general construction of divergence measures by representing the data's probability distribution as a point in the feature space~\cite{GreBorRasSchSmo07,Smola2007a,Song2010a,Xu2008}.

The idea of probability embedding is very simple.
Use a kernel to map the samples to the feature space, and take the mean to represent the (empirical) probability distribution.
This is possible because the feature space is a (linear) Hilbert space.
As long as one uses a powerful enough kernel, this simple process gives a unique representation of the empirical probability law that asymptotically converges to the true distribution.
Technically, it is sufficient to show that the kernel $\kernel$ is \textit{strictly positive definite} (spd) in the integral sense, that is,
\begin{align}
    \iint \kernel(x, y) p(x) p(y) \dm{x} \dm{y} > 0
\end{align}
for any probability distribution $p$ of consideration on the input space~\cite{Park2012a}.

Interestingly, the mean of the point process in the Hilbert space for the binned or linear functional kernel results in estimators for the firing rate function.
The PSTH can be formed by using a binned spike train kernel, and a smoothed estimate of the intensity function can be produced by using a linear functional kernel.
Given a collection of spike trains (with possible repeats), $\{x_i\}_i^N$, the mean in the Hilbert space corresponding to a linear functional kernel $\kernel$ can be represented as a function over arbitrary spike train $z$,
\[
    \frac{1}{N} \sum_i^N \kernel(x_i, z) = \frac{1}{N} \sum_i^N \sum_j^{N_i} \sum_k^{N_z} g(t^i_j, t^z_k).
\]
Since the kernel is linear for superposition, the mean does not depend on which spike train a particular spike came from. Therefore, the mean does not capture any statistical structure between spikes within the same trial.
As can be expected, neither the binned kernel nor the linear functional kernels is spd~\cite{Park2012a}.

For spd kernels, the mean contains all information about the collection of spike trains.
This is not surprising, given that unique spike trains mapped to the Hilbert space are not only unique, but are mutually linearly independent (the Gram matrix is full rank).
The mean ``remembers'' the set of spike trains that formed it, except for the ordering.
What is important is that the mean in the Hilbert space is a smoothed representation, and hence if the spike trains that consist the mean are similar, they are close in the Hilbert space.

A divergence measure for empirical observations can be defined as the distance of the means for a pair of collections of observed spike trains in the Hilbert space,
\begin{align}
    &\divv\!\left(\{x_i\}_i^{N}, \{y_j\}_j^{M}\right)^2
    = 
	\norm{
	    \frac{1}{N}\!\sum_i \kernel(x_i, \cdot)
	    -
	    \frac{1}{M}\!\sum_j \kernel(y_j, \cdot)
	}^2
    \nonumber \\
    &\quad=
	\frac{1}{N^2}\sum_{i,i'} \kernel(x_i, x_{i'})
	+
	\frac{1}{M^2}\sum_{j,j'} \kernel(y_j, x_{j'})
    \nonumber \\
    &\qquad
	-
	\frac{2}{N M}\sum_{i,j} \kernel(x_i, y_j)
\end{align}
where $\kernel$ is a spd kernel.
This divergence statistic $\divv$ is called \textit{maximum mean discrepancy} (MMD)~\cite{GreBorRasSchSmo07}.
When MMD is large, it is an indication that the two point processes are different.
In the classical hypothesis testing framework, we need the distribution of MMD under the null hypothesis which assumes that both collections originate from the same underlying random process.
We can generate MMD values from the null distribution by mixing the samples from both conditions and resampling from the mixture~\cite{GreBorRasSchSmo07,Park2012a}.
The following simple procedure describes a typical hypothesis testing given
two collections of spike trains $(x_i)_{i=1}^N$ and $(x_j)_{j=N+1}^{M+N}$, and a test size $\alpha$:
\begin{enumerate}
    \item Compute the kernel matrix $\kernelMatrix$
    \item Compute $ \divv^2\!
 	= \frac{1}{N^2} \ones\trp \kernelMatrix(I, I) \ones
 	+ \frac{1}{M^2} \ones\trp \kernelMatrix(J, J) \ones
 	- \frac{2}{N M} \ones\trp \kernelMatrix(I, J) \ones
    $
    \item Bootstrap randomly permuted indices of size $N$ and $M$ with replacement and recompute the statistic of the null distribution
    \item If $\divv^2$ is above the $(1-\alpha)$ quantile of the bootstrapped null distribution, reject the null hypothesis, otherwise accept it.
\end{enumerate}

The smoothness of the probability embedding is controlled by the spike train kernel of choice, and hence it is important to choose a kernel that captures the natural similarity of spike trains well.
This may come as a surprise since all spd kernel are asymptotically equivalent for MMD, that is, if the two underlying probability laws are different, any spd kernel can discriminate given a large enough sample.
Yet, the small sample power of the divergence test is greatly enhanced by encoding more prior information of the similarity into the spike train kernel.

\subsection{Neural decoding}

Neural decoding searches for the detailed relationship between neural activity and the variable of interest (Fig.~\ref{fig:decoding}).
Successful decoding analysis often provides evidence (or new hypothesis) for specific coding schemes the neural system uses, functionally identifies the system, and moreover, it can be used to develop neural prosthetics and interfaces.

Depending on the modality of the target variable, neural decoding can be tackled by different tools.
When the target variable is categorical (finite number of possibilities), classification algorithms are suitable; e.g., object recognition and multiple choice decision are naturally categorical.
If the target variable is continuous valued and fixed for each trial, but jumps from one value to another, then regression tools are appropriate.
Such trial-based experimental design is commonly used for studying the neural code.
When a continuous target variable is varied through time, filtering and smoothing algorithms are appropriate.
Most primary sensory as well as motor features naturally fall in this category, and are of most practical use in neural engineering.
Here we will focus our discussion on regression and filtering.
By regression, we mean batch analysis, while by filtering, we refer to online (real-time) signal processing.

A traditional approach to map single or multiple spike trains to a continuous target variable is linear regression on binned spike trains with relatively large bin sizes~\cite{Salinas1994,Churchland2001,Graf2011}.
Again, the rational stems from the neuroscience literature which focuses primarily on the information carried by the mean firing rate, and little about the detailed temporal structure within each trial.
Despite their crudeness, linear models on binned spike trains perform reasonably for motor brain machine interfaces, 
because the time scale of behavior is at the hundred milliseconds scale.

For filtering, conventional linear filtering methods such as least mean squares, recursive least squares, and Kalman filters are often used, and recurrent neural network approach for filtering is also worth mentioning~\cite{Sanchez2005}.
In recent years, state-space based Bayesian filtering approaches have been popular~\cite{Brown2001,Truccolo2005,Brown2004,arp:wang:09}.
A state-space (also known as latent-variable) model combined with an encoding model from continuous observation to spike trains is inverted using the Bayesian filtering framework.
This method requires a good encoding model which has to be fit ahead of time, and is based on stationary assumptions during and between training and testing conditions. 
Because of neural plasticity, in practice frequent retraining or sophisticated tracking is needed.

\subsubsection{Gaussian process regression}
\begin{figure}[t]
    \centering
    \includegraphics[width=0.48\textwidth]{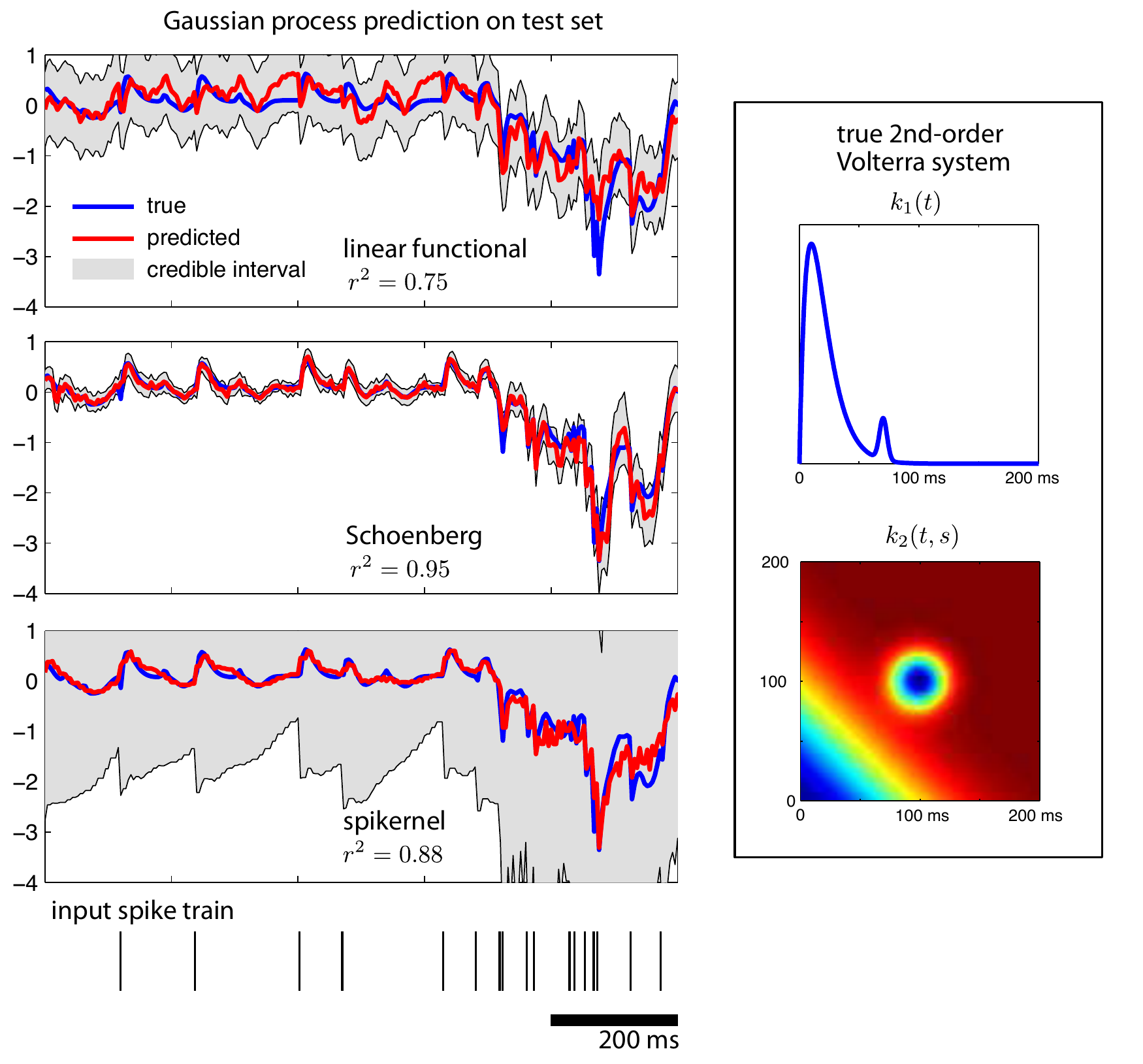}
\caption{
    Regression example using Gaussian processes on a synthetic problem.
    Gaussian process regression is used to estimate a target variable on a training set in a test set.
    Decoding traces with their credible intervals (2 standard deviation of the posterior around the mode) are plotted. 
    The target variable is generated through a second-order Volterra system given by,
    $y(t) = 0.1 + \sum_i k_1(t - t_i) + \sum_{i,j} k_2(t_i, t_j)$ (right).
    (Volterra kernels $k_1$ and $k_2$ are not to be confused with the spike train kernel used for decoding.)
}
\label{fig:gp}
\end{figure}

Gaussian process (GP) regression is a widely used nonparametric Bayesian regression method~\cite{Rasmussen2005}.
For neural decoding, we assume a prior distribution over the functions from spike trains to $\reals$.
This prior is explicitly stated to be a Gaussian process with covariance provided by a kernel;
the covariance of the function values evaluated at two spike trains $x$ and $y$ is given by 
$\cov(f(x), f(y)) = \kernel(x, y)$.
To be specific, given a set of spike trains $\{x_i\}_{i=1}^n$, the distribution of the function values from the prior is multivariate Gaussian distributed
\[
    [f(x_1), \ldots, f(x_n)] \sim \mathcal{N}(0, \kernelMatrix)
\]
where $\kernelMatrix$ is the kernel matrix~\eqref{eq:def:kernel:matrix}.

Using the GP prior, we can infer the posterior assuming a Gaussian noise corrupted observation model.
The prediction of the function evaluated at spike train $z$ is given by,
\begin{align}\label{eq:GP:prediction}
    \hat f(z) = \sum_i^n \alpha_i \kernel(x_i, z)
\end{align}
where $\mathbf{\alpha} = (\kernelMatrix + \sigma_n^2 I)^{-1} \vy$, 
where $\sigma_n^2$ is the observation noise variance and $\vy$ is the desired signal vector corresponding to training data $\{x_i\}_{i=1}^n$.

There are several advantages of GP:
\begin{enumerate}
    \item the prediction coincides with kernel ridge regression (regularized kernel least squares), but GP provides the posterior credible interval (not to be confused with the frequentist confidence interval) which indicates the uncertainty of the prediction under the model assumptions,
    \item given a universal kernel, it can learn any nonlinear functional relation, and,
    \item hyperparameters (kernel parameters and observation noise variance) can be tuned in a principled way using empirical Bayes procedure.
\end{enumerate}

In figure~\ref{fig:gp}, we compare GP regression with linear functional, Schoenberg, and spikernel in a synthetic example where a Poisson spike train is mapped to a real-valued signal through a second order Volterra system.
The hyperparameters are learned through empirical Bayes method where the marginal likelihood is maximized on the training set (400 points).
The linear functional kernel of~\eqref{eq:mci} does not perform well on mean prediction (red trace) because of the strong nonlinear component (pairwise interaction of spike timings due to the second order Volterra kernel), while the spikernel obtains a reasonable prediction, and Schoenberg kernel of~\eqref{eq:schoenberg} achieves very high performance.
The credible interval resulting from using the Schoenberg kernel is the smallest, meaning the model is confident that the data is well described by the regression result.
In contrast, the inferred credible interval for the spikernel is large, meaning at least some aspects of the data are not well described by the fit model.

\subsubsection{Kernel adaptive filtering}

For closed loop applications, the system identification and prediction benefit from sequential processing where the system parameters are adapted with every new sample because neural systems are plastic and there are real time constraints in the experimental setup.
Therefore, adaptive filtering algorithms have been widely used in the brain-machine interface applications, and other neural prosthetics~\cite{Nicolelis2009}.
As stated, linear filtering algorithms such as least mean squares (LMS) and recursive least squares (RLS) algorithms as well as Kalman filtering have been successful using the binned representation, but performance improvements are still needed.

Kernel adaptive filters (KAF) have been recently developed that kernelize the linear adaptive filtering algorithms~\cite{Liu2010}, inheriting their simple computational structure, and extending them to nonlinear transfer functions.
Similarly, KAFs operate on a sample-by-sample basis, and can deal with non-stationary environments. 

\begin{figure*}[t!]
\centering
\includegraphics[width=\textwidth]{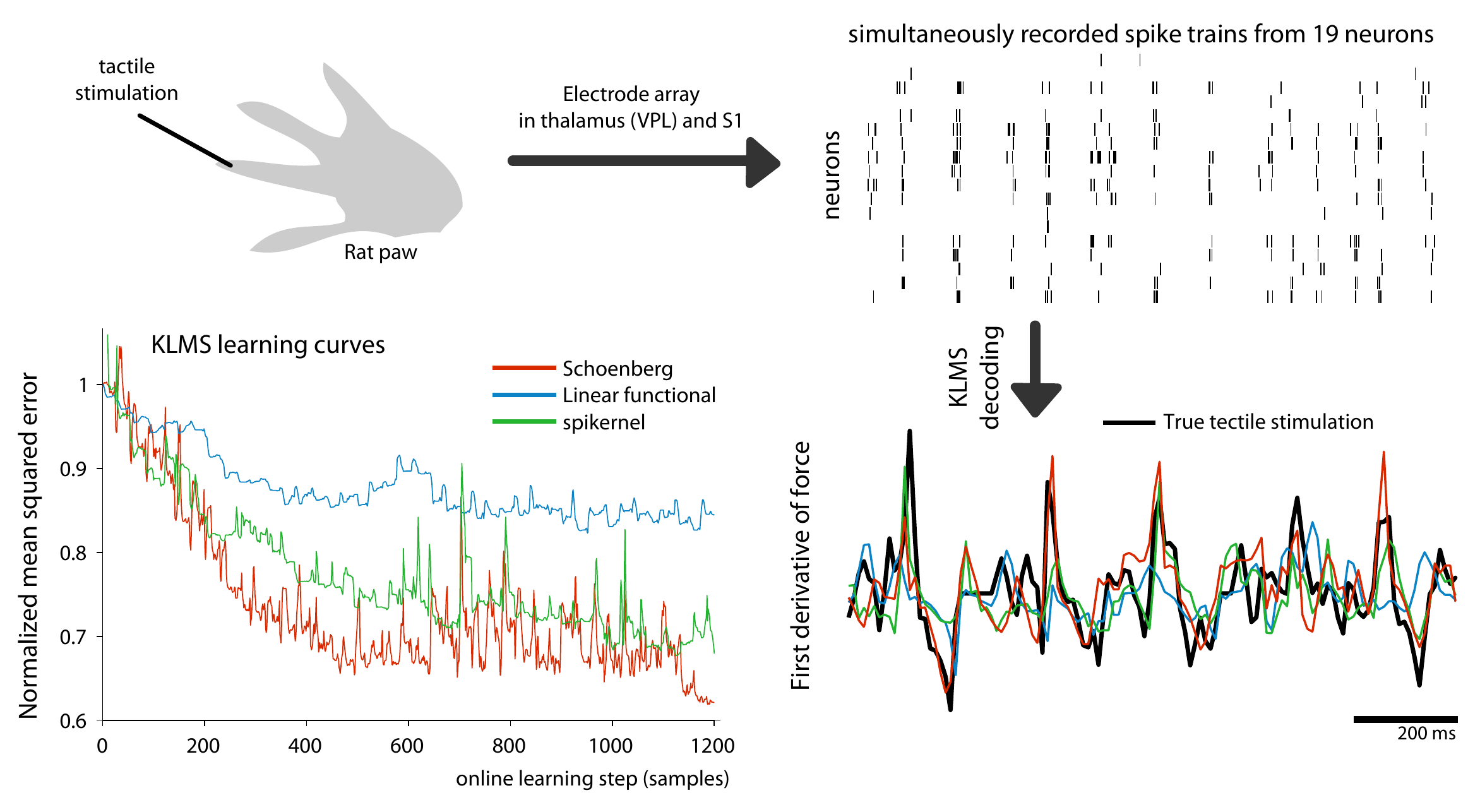}
\caption{
    Decoding tactile stimulus from awake rat thalamus and somatosensory spike trains using kernel least means square (KLMS) algorithm.
    80 ms sliding window over the spike trains is used to decode stimulus sampled at 100 Hz.
    Hyperparameters (kernel size and learning rate) are optimized for best performance on a cross-validation data (1 sec).
    First derivative of the force applied to the finger is the target variable, and 19 neurons. 
    Learning curve shows the convergence of online learning (12 seconds of training is shown) quantified on a test dataset of 2 seconds.
    (Adapted from~\cite{Li2012b})
}
\label{fig:KLMS}
\end{figure*}

Here we describe the simplest yet powerful kernel least mean squares (KLMS) algorithm, which has been successfully applied in neural engineering as inverse control in the context of somatosensory prosthetics~\cite{Li2012a}.
KLMS is a nonlinear extension of the popular LMS algorithm that can be derived as a stochastic gradient descent on the mean squared error but with an infinite dimensional weight vector.
The filter mapping is a function of the same form as \eqref{eq:GP:prediction}, where the $\alpha$'s are proportional to the instantaneous errors:
\begin{align}
    \alpha_i &= -\eta (y_i - \hat{f}_{i-1}(x_i))
\end{align}
where $\hat{f}_{i-1}$ is the estimated filter before the $i$-th observation, and $|\eta| < 1$ is the learning rate~\cite{Liu2010}.
Thus, the filter input-output map is fully represented by pairs of real coefficient $\alpha_i$ and observed spike train $x_i$.
The KLMS has been applied with advantages in nonlinear signal processing of continuous amplitude signals, mostly using the Gaussian kernel, but one of the advantages of the RKHS approach is that the algorithm formulation is independent of the kernel. 
Therefore, any of the spike train kernels presented in this paper can be directly used in the KLMS.

We demonstrate online inverse modeling using KLMS in figure~\ref{fig:KLMS}~\cite{Li2012b,Li2012phd}.
The goal is to reconstruct properties of the induced tactile stimulation (here the derivative of the force) from the generated multi-channel spikes in the thalamus (VPL) and primary sensory areas (S1) upon stimulation.
Given the finite memory in the neural system, typically a moving window is used—for example, of 80 milliseconds slided 2 ms at a time over the spike data.
We use the product kernel in this MISO KLMS model, train a decoder on the tactile stimulus time series on a training set, and compare reconstruction performance with different spike train kernels on a test set.
Overall, a similar trend to previous examples is observed in this application where the Schoenberg kernel outperforms the spikernel and the linear kernel, both in terms of faster convergence in the training set and better reconstruction of the stimulus in the test set. 

\section{Discussion}
Spike train kernels enable signal processing and machine learning of spike trains by providing a feature space for computation.
In this article, we surveyed how positive definite functions can be used to quantify spike trains, at the rate or spike timing level, and implement important operations for neural decoding such as hypothesis testing using probability embedding, and estimation of continuous functional mappings from spike trains.
As we have briefly demonstrated, spike train kernels provide a unifying framework for most of the techniques used in spike train processing, from the statistical rate descriptors, to the binned representations up to the full description (injective mapping by a spd kernel) of the spike train timing structure in infinitely dimensional functional spaces.
The approach is therefore versatile and mathematically principled, extending the footprint of signal processing and machine learning to more abstract spaces. 

Among the spike train kernels, we have promoted the binless Schoenberg kernel, since, (1) it provides an injective mapping, (2) it can embed arbitrary stochasticity of neural responses as the sample mean in the RKHS, and (3) it is a universal kernel that can approximate arbitrary functions on spike trains.
Such arbitrarinesses in the point process and function classes are allowed because of the strongly nonparametric nature (spd) of the kernel mapping.

\subsection{Interpretation of decoding analysis}
Like other advanced signal processing methodologies (e.g., deep learning and nonparametric Bayesian methods), strictly positive definite spike train kernels make the results less interpretable due to the high dimensionality of the implicit feature space.
Weaker kernels only encapsulate explicitly certain designed features, for instance, the count kernel is only sensitive to the total spike count, and the linear functional kernel is mostly sensitive to firing rate profiles.
Although the stronger kernels can capture arbitrary features, they are not unique.
Therefore designing explicitly stronger spike train kernels is non-trivial, because it is hard to understand what spike train features they are emphasizing.
There are several ways we can partially recover the intuition, although more research is needed in this direction:
via visualization of spike trains in the feature space in the case of MMD-PCA (Fig.~\ref{fig:divergence}), via sparse regression methods like relevant vector machine, or via kernel selection over a set of strongly interpretable weaker (more parametric) kernels.

Another caveat of decoding analysis is that successful decoding does not imply that the brain is using the information, it only signifies that the information exists in the collected neural signals.
For example, 
in early sensory neurons like retinal ganglion cells or auditory fiber,
we can often decode and reconstruct the sensory stimulus better than what the subject can report by behavior.
Therefore, we should be cautious not to over-interpret successful decoding.

\subsection{Future directions}
The last decade has been productive in terms of new kernels for many abstract (non-Euclidean) spaces, but there is still room for improvement.
We would like to have a spike train kernel that is powerful enough, i.e.,  spd and universal, while at the same time, able to capture our prior knowledge about the similarity between spike trains.
The variability of the neural system could provide hints to designing better spike train kernels.

There are three practical aspects of designing a useful kernel:
(1) the kernel should encode the prior knowledge of the input domain, and the problem at hand, 
(2) the kernel should be computationally tractable, perhaps with linear or less time complexity in the number of spikes; and 
(3) the kernel should have no or very few parameters---in the latter case there should be simple intuition behind the parameters, and more importantly simple rules for setting their values.

We have discussed two frameworks in this article, namely the binned kernels, and the functional kernels.
Binned kernels are either too simplistic, they ignore the temporal structure, or computationally too expensive, e.g., spikernel.
On the other hand, some functional kernels are either overly sensitive to the mean rate, such as linear functional kernel, or involve parameters that are not easily visualized.
A kernel with the right balance among these three properties remains to be found.
It is safe to assume that we have only scratched the surface of this problem, and there remain many open avenues to be explored.
Two possible approaches are edit kernels, and generative kernels.
Edit kernels rely on the principle of adopting simple operations such as shifting, addition, and deletion to convert one spike train into another where each operation is assigned a cost, whereas generative kernels rely on the principle that two spike trains are similar if they originate from the same generative model.

\bibliographystyle{IEEEtran}

\end{document}